\documentstyle[aps,prl,twocolumn,epsfig,floats]{revtex}

\newcommand{\equ}[1]{(\protect\ref{#1})}
\newcommand{\jam}{\theta_\infty}
\newcommand{\avern}{\left< n \right>}
\newcommand{\reals}{{\rm I \mkern-2.5mu \nonscript\mkern-.5mu R}}
\newcommand{\Journal}[4]{#1 {\bf #2}, #3 (#4)}

\begin{document} 

\draft
\wideabs{

\title{Long-range-interactions induced ordered structures 
in deposition processes} 

\author{R. Pastor-Satorras}
\address{Department of Earth, Atmospheric, and Planetary Sciences\\
Massachusetts Institute of Technology, Cambridge, Massachusetts
02139, USA}
\author{J. M. Rub\'{\i}}
\address{Departament de F\'{\i}sica Fonamental, Facultat de
F\'{\i}sica\\ 
Universitat de Barcelona, Diagonal 647, 08028 Barcelona, Spain}

\maketitle

\begin{abstract}
  We present a new model of sequential adsorption in which the
  adsorbing particles experience dipolar interactions.  We show that
  in the presence of these long-range interactions, highly ordered
  structures in the adsorbed layer may be induced at low temperatures.
  The new phenomenology manifests through significant variations of
  the pair correlation function and the jamming limit, with respect to
  the case of noninteracting particles.  Our study could be relevant
  in understanding the adsorption of magnetic colloidal particles in
  presence of a magnetic field.
\end{abstract}


}

The study of the irreversible adsorption of colloidal particles onto a
surface has been since long time subject of a great deal of interest,
due to its potential applications to physical, physico-chemical, and
biological problems \cite{evans93}.  Our global understanding of the
process has been possible through the formulation of different models,
analyzed either numerically or analytically.  These models share a
sequential and irreversible nature, and differ in the rules by which
the particles accommodate when arriving at the surface. The various
rules are responsible for the different values of the relevant
quantities describing the adsorbed phase, such as the jamming limit,
the pair correlation function, and the local variance of the number
of deposited particles.  In the random sequential adsorption (RSA)
model \cite{renyi58,feder80,schaaf88,senger91,ramsden93} particles are
placed at random positions on the substrate. If an incoming particle
overlaps with a previously adsorbed one, it is rejected and a new one
probed; otherwise it becomes irreversibly adsorbed. In the ballistic
model (BM) \cite{meakin87,talbot92,jullien92,thompson92} the particles
descend to the surface following straight vertical trajectories. An
incoming particle that does not reach the substrate is allowed to roll
over the previously adsorbed, following the steepest descent path,
until it reaches a stable position. Only particles that fail to gain
the surface are finally rejected.

All these models, and their subsequent extensions, have been mainly
implemented by considering short range---hard core---interactions among
particles.  With the exception of the analysis of the role played by
electrostatic interactions \cite{adamczyk96}, the case of long-range
interactions remains essentially unexplored.  Our purpose in this
Letter is to analyze comprehensively the influence that these
interactions have in the kinetics of deposition in a simple numerical
model.  We will show that when they are taken into account, a new
aspect of the problem emerges. The structure of the adsorbed layer
changes considerably, giving rise in some cases to the appearance of a
higher degree of order in the substrate.

To illustrate this point, we will focus on the case of anisotropic
dipolar interactions, and present numerical simulations of the
adsorption process on a line $(1+1d)$ and on a plane $(1+2d)$. In our
simulations we consider the adsorption of spherical magnetic particles
of diameter $a$ and magnetic moment $\vec{\mu} = \mu \vec{u}$, with
$\mu$ being the magnetic moment strength and $\vec{u}$ a unit vector
oriented along its direction. Orientation of dipoles is restricted to
the space in which they diffuse: $\reals^2$ in $(1+1d)$ and $\reals^3$
in $(1+2d)$.  The dipolar interaction between two particles $i$ and
$j$, located at positions $\vec{r}_i$ and $\vec{r}_j$, respectively,
is $U_{ij}= \mu^2 E_{ij}/a^3$, with $E_{ij}$ being the dimensionless
energy
\begin{equation}
  E_{ij}= a^3 \left\{ \vec{u}_i \cdot \vec{u}_j - 3(\vec{u}_i \cdot
    \vec{r}_{ij}) (\vec{u}_j \cdot \vec{r}_{ij}) / r_{ij}^2 \right\} /
  r_{ij}^3, 
\label{energy}
\end{equation}
where $\vec{r}_{ij} = \vec{r}_i - \vec{r}_j$.  Particles are released
at random positions over a certain initial height $z_{in}$, having
assigned an initial orientation $\vec{u}_i^0$.  Each particle
undergoes a random walk until it either becomes adsorbed or moves away
from the substrate a distance greater than $z_{out}$; in this case,
the particle is removed and a new one released.  The effect of the
interactions is taken into account by means of a Metropolis algorithm
\cite{binder86,pastor95,pastor98}. Suppose there are $N-1$ adsorbed
particles, located at points $\vec{r}_i$, $i=1,\ldots,N-1$.  At some
time $t$ an incoming dipole occupies the position $\vec{r}_N$ and has
a total energy $E=\sum_{i=1}^{N-1} E_{Ni}$.  At time $t+1$ we compute
a new position ${\vec{r}_N}^{\:*}$; the particle arrives there by
means of jump of length $\delta$ in a direction chosen at random. The
movement to ${\vec{r}_N}^{\:*}$ is performed rigidly, without changing
$\vec{u}_N$.  The energy experienced in the new position is $E^*$, and
the total change in the energy due to the movement is $\Delta E = E^*
- E$. If $\Delta E \leq 0$, the movement is accepted; otherwise, it is
accepted with probability $p=\exp(-\Delta E / T_r)$, where $T_r=a^3
k_{\mbox{\rm\scriptsize B}} T / \mu^2$ is a reduced temperature.  Note
that with this procedure we do not take into account gravitational
forces.  In order to speed up the algorithm, when an incoming particle
reaches a position very close to the substrate (less than one
diameter) it is attached according to the BM rules.  This procedure
indeed assumes the presence of a strong short-ranged attraction
between the substrate and the particles, which overcomes dipolar
interactions at short distances.

\begin{figure}[t]
  \centerline{\epsfig{file=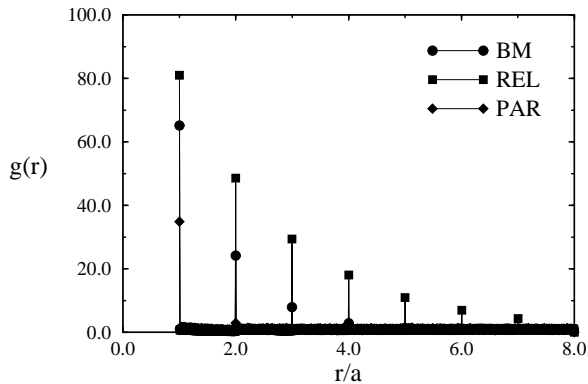, width=8cm}}
  \caption{Pair correlation function for dipolar adsorption, in the
    free limit $T_r\to\infty$, and for models REL and PAR at $T_r=0$.}
  \label{Correls}
\end{figure}

\begin{figure}[t]
  \centerline{\epsfig{file=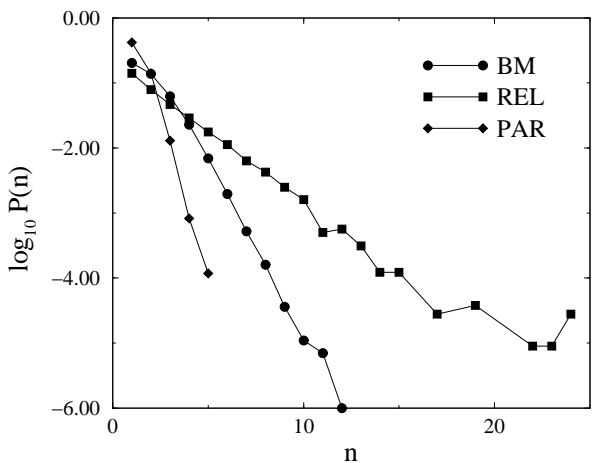, width=8cm}}
  \caption{Chain-length density function for dipolar adsorption, in
    the free limit $T_r\to\infty$, and for models REL and PAR at
    $T_r=0$.}
  \label{Size}
\end{figure}

The description of the algorithm is completed by prescribing the
dynamics of the dipoles. We have considered two possibilities under
scrutiny, defining two different models: (i) Relaxed dipoles (model
REL). The initial orientation $\vec{u}_i^0$ of the dipoles is assigned
at random. After every accepted movement, the moment of the random
walker is oriented along the direction of the total magnetic field on
its position.  After adsorption, a newly attached dipole undergoes a
last relaxation, and its direction does not change any more.  There
are two natural choices for the orientation of the first adsorbed
dipole: either select it randomly or parallel to the substrate. Both
cases provide comparable results, within the error bars. The results
reported here correspond to a random initial dipole.  (ii) Parallel
dipoles (model PAR). All dipoles are parallel to the substrate,
pointing rigidly towards a predetermined fixed direction.  Both models
are actually the limits of a process of adsorption of relaxing dipoles
in the presence of a magnetic field. Model REL is recovered in the
limit of zero field, whereas model PAR corresponds to a very strong
field.

In $(1+1d)$ we have simulated the adsorption of dipoles at different
temperatures onto a substrate of length $L$. The results reported here
are mainly for $L=150a$; larger system sizes provided systematically
equivalent results.  We have selected the parameters $\delta=a/2$,
$z_{in}=15a$, and $z_{out}=20a$. These are conservative estimates;
since dipolar interactions decay in average as $r^{-3}$, the energy
diminishes more than $3$ orders of magnitude from the neighborhood of
the substrate to the launching height $z_{in}$. Higher values of
$z_{in}$ and $z_{out}$ were tested, providing comparable results but
dramatically increasing the computation time.

The presence of dipolar interactions is expected to induce changes in
the structure of the adsorbed phase. We have investigated this point
in $(1+1d)$ by computing the density at jamming $\jam$, the pair
correlation function $g(r)$, and a new quantity, the {\em chain-length
density function\/} $P(n)$.  The BM rule adopted for the final
allocation of the dipoles allows an incoming particle to roll over
another, adsorbing therefore close together and forming chains---sets
of contiguous particles---in $(1+1d)$ \cite{notechain}. We define
$P(n)$ as the density of chains of length $n$, per unit length of
substrate. This function is related to the jamming limit through $\jam
= \sum_{n=1}^\infty n P(n)$.

In order to first check our algorithm, we have generated 1000 jammed
configurations on a linear substrate of $L=750a$, in the limit
$T_r\to\infty$ (free BM). We recover a jamming limit
$\jam^{BM}=0.8079\pm0.0004$, in excellent agreement with theoretical
predictions $\jam^{th}\approx0.80865$ \cite{talbot92}. The REL and PAR
models render in the most extreme case of $T_r=0$ (limit of zero
temperature or infinitely strong interactions) the jamming limits
$\jam^{REL}=0.855\pm0.002$ and $\jam^{PAR}=0.752\pm0.002$,
respectively.  The difference with the BM value is rather notable,
larger than a $5.7\%$ and $7.0\%$, respectively.  This feature should
be contrasted with the small variations in the jamming limit due to
the consideration of hydrodynamic interactions in the process (around
$1.4\%$) \cite{pago94}.

Figure \ref{Correls} depicts the pair correlation function $g(r)$ for
BM, and both models REL and PAR at $T_r=0$. Model REL enhances greatly
the maxima of $g(r)$, which occur at distances $r_p = ap$, for integer
$p$ (that is, corresponding to interparticle distances equal to a
multiple of the diameter). We find in this case that the peaks decay
exponentially, defining a correlation length of the form $g^{REL}(r_p)
\sim \exp(-r_p/\xi)$.  We estimate $\xi=4.73\pm0.04$. In model
PAR, on the other hand, correlations are strongly suppressed; the
peaks in the function are almost invisible for $r>2a$.

\begin{figure}[t]
  \centerline{\epsfig{file=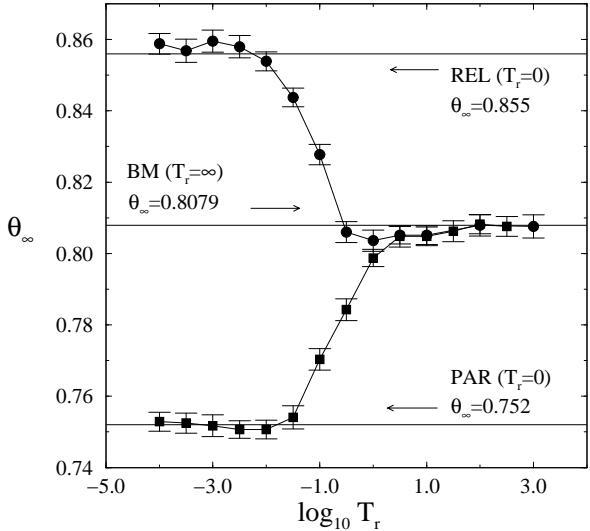, width=8cm}}
  \caption{Jamming limit $\jam$ as a function of the reduced temperature
    $T_r$ for the models REL (circles) and PAR (squares).}
  \label{Jam}
\end{figure}

Given the expression \equ{energy} for the dipolar energy, it is
energetically favorable for the particles to form linear chains of
aligned dipoles. This effect is studied in Fig.~\ref{Size}, where we
have plotted $P(n)$ for BM, and both REL and PAR  at $T_r=0$.  We
observe again an exponential decay for REL, $P^{REL}(n) \sim
\exp(-n/\bar{n})$, with an estimated correlation size $\bar{n} =
4.67\pm0.05$, in good agreement with the characteristic length for the
correlation function, $\xi=4.73$.  This exponential behaviour, on the
other hand, is lost in BM and PAR; in those cases, the function $P(n)$
decays much faster.  The most characteristic property of
Fig.~\ref{Size} is the way in which the weight of the distribution
moves towards larger chains in model REL (up to $n=24$, which is
almost 1/6 of the total substrate length). Large chains are, on the
contrary, strongly suppressed in model PAR (only $n=5$, 1/30 of the
substrate length). The witdh of the chain-length distribution can be
estimated by means of the average chain length $\avern$,
\begin{equation}
\avern = \frac{\sum_{n=1}^\infty n P(n)}{\sum_{n=1}^\infty P(n)}.
\end{equation}
We observe the values $\avern^{BM}=1.857\pm0.003$,
$\avern^{PAR}=1.298\pm0.007$,  and $\avern^{REL}=2.52\pm0.02$.

\begin{figure}[t]
  \centerline{\epsfig{file=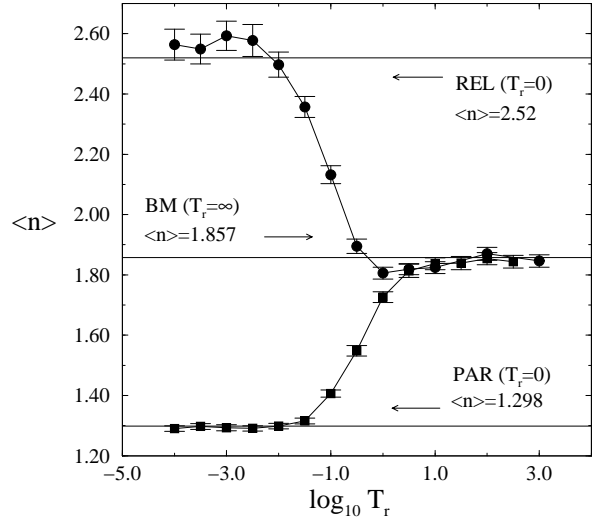, width=8cm}}
  \caption{Average chain length $\avern$  as a function of the reduced
    temperature 
    $T_r$ for the models REL (circles) and PAR (squares).}
\label{Aver}
\end{figure}

The next point we have analyzed is the dependence of the relevant
quantities of the models on the temperature.  In Figure~\ref{Jam} we
have plotted the jamming limit $\jam$ for models REL and PAR, as a
function of the reduced temperature $T_r$. A similar plot is made in
Fig.~\ref{Aver} for the average chain length $\avern$. From these two
figures we conclude that, when increasing $T_r$, $\jam$ and $\avern$
jump rather discontinuously towards the limit value at high
temperature (BM).  This behaviour contrasts with the results reported
for the diffusion-limited aggregation (DLA) of particles with dipolar
interactions \cite{pastor95,pastor98}. In those works, the fractal
dimension of the agregates was computed as a function of a reduced
temperature $T_r$. It was found that the fractal dimension of dipolar
DLA also changes between a high and a low temperature values, with a
transition that seems to be continuous, taking place during almost 4
orders of magnitude in $T_r$. In the present case, however, we observe
a transition ocurring in a range of about 1.5 orders of magnitud in
$T_r$. It seems therefore more likely the presence of a crossover in
dipolar adsorption. The discontinuous nature of this crossover is
smoothed by finite-size effects, but would be more evident in larger
systems.  We note that the crossover would happen at different
characteristic temperatures $T_c$ for the two models. In view of
Figs.~\ref{Jam} and~\ref{Aver}, we estimate the values $T^{REL}_c
\approx 10^{-1.25}$ and $T^{PAR}_c \approx 10^{-0.5}$.

\begin{figure}[t]
  \centerline{\epsfig{file=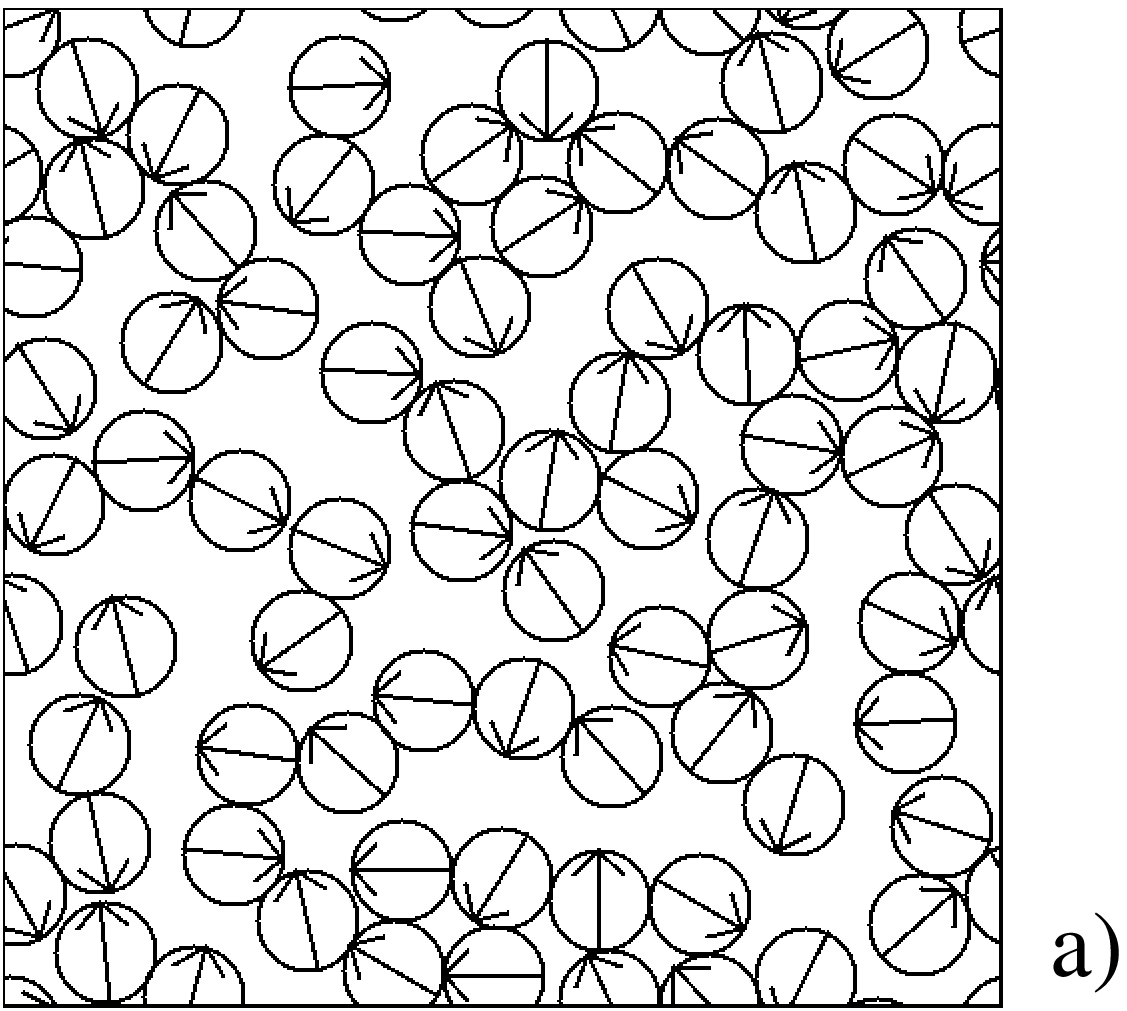, width=7cm}}
  \centerline{\epsfig{file=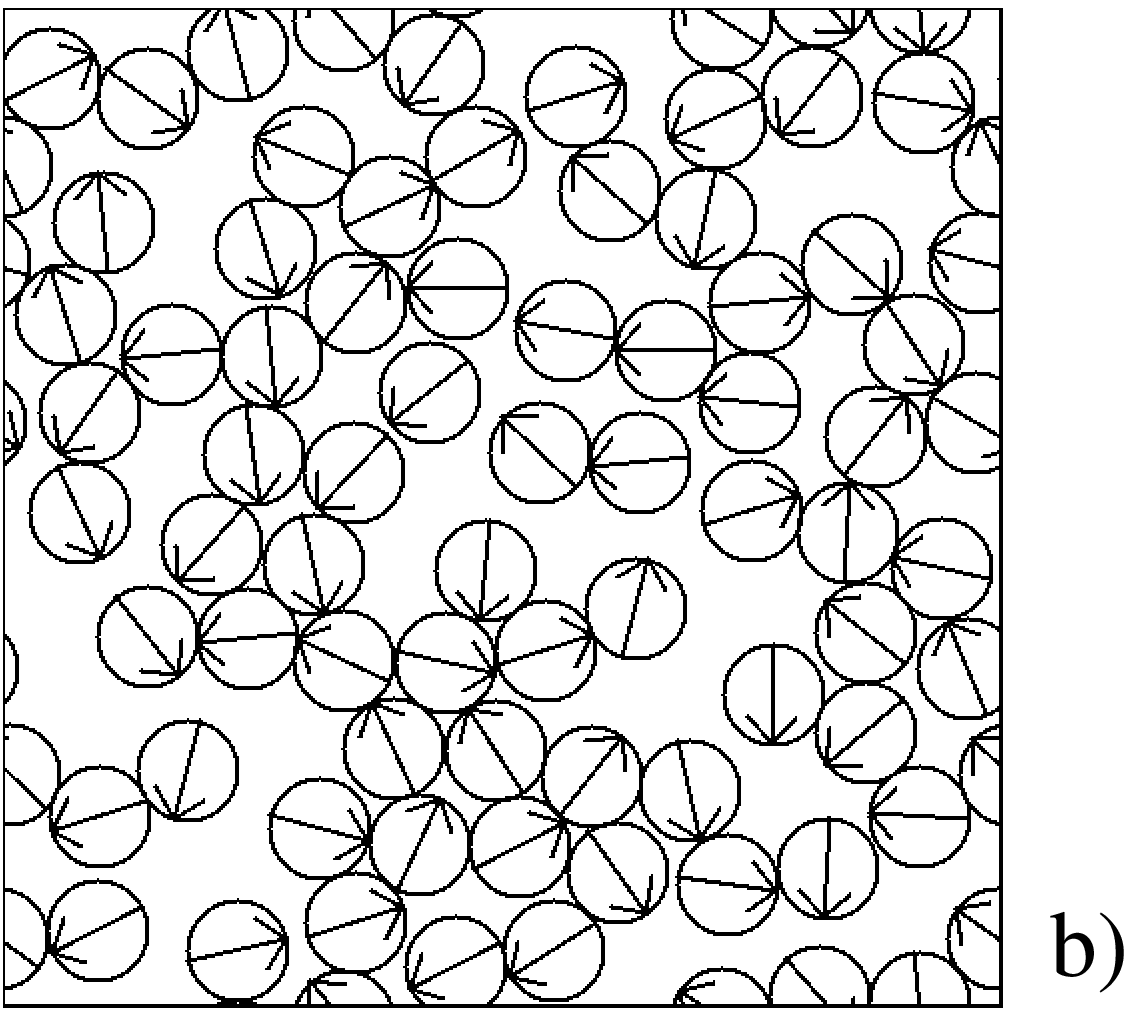, width=7cm}}
  \vspace*{0.25cm}
  \caption{Jammed configurations in a square of size $L=10a$. a)
    Free BM. b) REL model. In this last case one can appreciate the
    relative tendency of neighbor dipoles to orient parallel to each
    other.}
\label{Patterns2d}
\end{figure}

Finally, in order to study the effects of dimensionality, we have
considered the extension of our model to $(1+2d)$, by adsorbing
dipolar particles onto a plane of size $L \times L$. In the case of
free BM, we recover a jamming limit $\jam=0.609\pm0.003$, in good
agreement with previous simulations \cite{jullien92,thompson92}. When
introducing dipolar interactions at $T_r<\infty$, the algorithm turns
out to be extremely time consuming, specially for model PAR.  We
report here preliminary results only for model REL at $T_r=0$, on
surfaces of a fairly small size ($L=10a$). Figure \ref{Patterns2d}
depicts typical configurations in free BM and REL.  The arrows drawn
correspond to the projection onto the adsorption plane of the
originally tridimensional dipoles.  We can observe the local effect of
the interactions, ordering the dipoles of neighboring particles in
order to minimize the dipolar energy \equ{energy}. The estimated value
of the jamming limit is in the case REL $\jam=0.622\pm0.003$. The
difference with respect to the free case is $1.8\%$, smaller than that
observed in $(1+1d)$. This is a result to be expected, since
increasing the dimensionality increases the possibilities for disorder
in the system. Therefore, the relative packing of the final state is
expected to be smaller than in $(1+1d)$. A similar effect was also
observed in $3d$ simulations of dipolar DLA \cite{pastor95}.

In summary, we have shown that the presence of long range dipolar
interactions during the adsorption process of particles on a substrate
induces important changes in the structure of the adsorbed phase,
reflected in the jamming limit, correlation function, and average
chain length.  In $(1+1d)$, one of the two models considered, namely
relaxed dipoles (REL), shows long-range order, manifested in the
tendency of the particles to form long chains, increasing therefore
the packing of the substrate.  On its turn, the model of parallel
dipoles (PAR) produces smaller chains and results in a more disordered
jammed phase.  These results can be regarded as an effect of the
different dynamics of the dipoles.  Given the form of the dipolar
energy \equ{energy}, the adsorbed rigid dipoles in model PAR tend to
repel an incoming particle approaching normally the surface; this fact
induces an average repulsion between particles and substrate. On the
other hand, the relaxing dynamics in model REL results in an
induced attraction between dipoles and surface, attraction responsible
for the larger surface coverage.  When decreasing temperature, we 
observe in model REL a transition from a disordered phase at high
$T_r$ to an ordered phase emerging at low temperatures; in model PAR
the transition is reversed.  In the $(1+2d)$ geometry the effects are
less pronounced, reflected in a slight variation of the jamming limit
and an induced orientational order in the dipoles.  Our findings may
open further perspectives for the development of new models accounting
for the phenomena we have reported and the different properties of the
substrate, as well as a technical hint to achieve an increased
efficiency of the adsorption processes, allowing the possibility of
covering with particles a larger fraction of the surface.

We would like to thank Olav van Genabeek for many insights that helped
in coding the model. RPS benefited from a scholarship grant from
the Ministerio de Educaci\'{o}n y Cultura (Spain). JMR acknowledges
financial support by CICyT (Spain), Grant no. PB92-0895.


\begin{references}
\bibitem{notechain} 
  In $(1+2d)$ the particles can adsorb in multiply connected
  configurations. The notion of chain loses therefore its meaning.
\end{references}
\end{document}